\documentclass[conference]{IEEEtran}
\IEEEoverridecommandlockouts
\usepackage{cite}
\usepackage{amsmath,amssymb,amsfonts}
\usepackage{algorithmic}
\usepackage{graphicx}
\usepackage{textcomp}
\usepackage{xcolor}
\def\BibTeX{{\rm B\kern-.05em{\sc i\kern-.025em b}\kern-.08em
    T\kern-.1667em\lower.7ex\hbox{E}\kern-.125emX}}
\begin{document}

\title{Higher Order Temporal Analysis of Global Terrorism Data}

\author{\IEEEauthorblockN{Madelyn Dunning}
	\IEEEauthorblockA{\textit{Pacific Northwest National Laboratory} \\
		Richland, WA, USA \\
		Madelyn.Dunning@pnnl.gov}
	\and
	\IEEEauthorblockN{Sumit Purohit}
	\IEEEauthorblockA{\textit{Pacific Northwest National Laboratory} \\
		Richland, WA, USA \\
		Sumit.Purohit@pnnl.gov}
}

\maketitle

\begin{abstract}
Temporal networks are a fundamental and flexible way of describing the activities, relationships, and evolution of any complex system. Global terrorism is one of the biggest concerns of recent times. It is also an example of a temporal network that evolves over time. Graph analytics can be used to explore salient properties of the terrorism network to understand its modus operandi, which can be used by the global alliance of security and government entities to form a co-ordinated response to this threat. We present graph based analysis to understand temporal evolution of global terrorism using the Global Terrorism Database (GTD). 
\end{abstract}


\section{Introduction}
Networks are a fundamental and flexible way of representing entities, relationships, and behaviors in many real-world domains such as power grids \cite{chu2017complex}, social networks \cite{kumar2010structure}, modeling adversarial activities \cite{cottam2018multi}, and terrorist networks \cite{belli2014exploring}. Temporal evolution of such networks is of great interest to understand how a specific network changes over the course of time and whether can we predict the changes expected in the future. Graph theoretic metrics that are sufficient to model a static network fail to capture non-linear dynamic behavior of a temporal network. Global terrorism involves a network of terrorist organizations and their sympathizers, and has a long history of perpetrating attacks on the social, political, and economic stability of different regions of the world. We present a graph based approach to investigate the relationships and behavior of such organizations which cannot be captured using naive count-based tabular analysis. We model these organizations based on their involvement in a set of terrorist events. We use the Global Terrorism Database (GTD) \cite{lafree2007introducing} \cite{startgtd}, which is an open-source database including information on terrorist events around the world from 1970 through 2017. It contains information about 181,000 terrorist events, with a total of 135 attributes for each event, including date, location, group, weapon, casualty and so on. Some of the more useful variables in the GTD include: ID, Date, Location, Summary, Attack type, Target type (for example, Police Checkpoint), Target subtypes (for example, Iraqi Police Service), and specific targets, Group name if known, and Weapon types. When the attack is part of a coordinated multi part incident, then the related incident IDs are also listed. It is maintained by the National Consortium for the Study of Terrorism and Responses to Terrorism (START).

\section{Related Work}
Steve Ressler \cite{ressler2006social} presents a survey of social network analysis approaches to combat terrorism. It distinguishes terrorist organizations from hierarchical, state-sponsored appointments in characteristics such as leadership and organizational structure. It uses networks to analyze recruitment, evolution, and the diffusion of radical ideas. Fellman et al. \cite{fellman2008complexity} uses non-linear dynamical systems modeling to explore centrality and hierarchy of 9-11 hijackers network. Carley et al. \cite{carley2003destabilizing} presents dynamic network analysis to understand evolution of a network to destabilize a terrorist network. Our work presents a real-world use case and uses temporal graph patterns to model evolution of a terrorist network. We present tools to analyze large scale graphs.

\section{Approach}
\subsection{Static Graph Analysis}
\begin{figure}
	\centering
	\includegraphics[width=280pt,height=300pt]{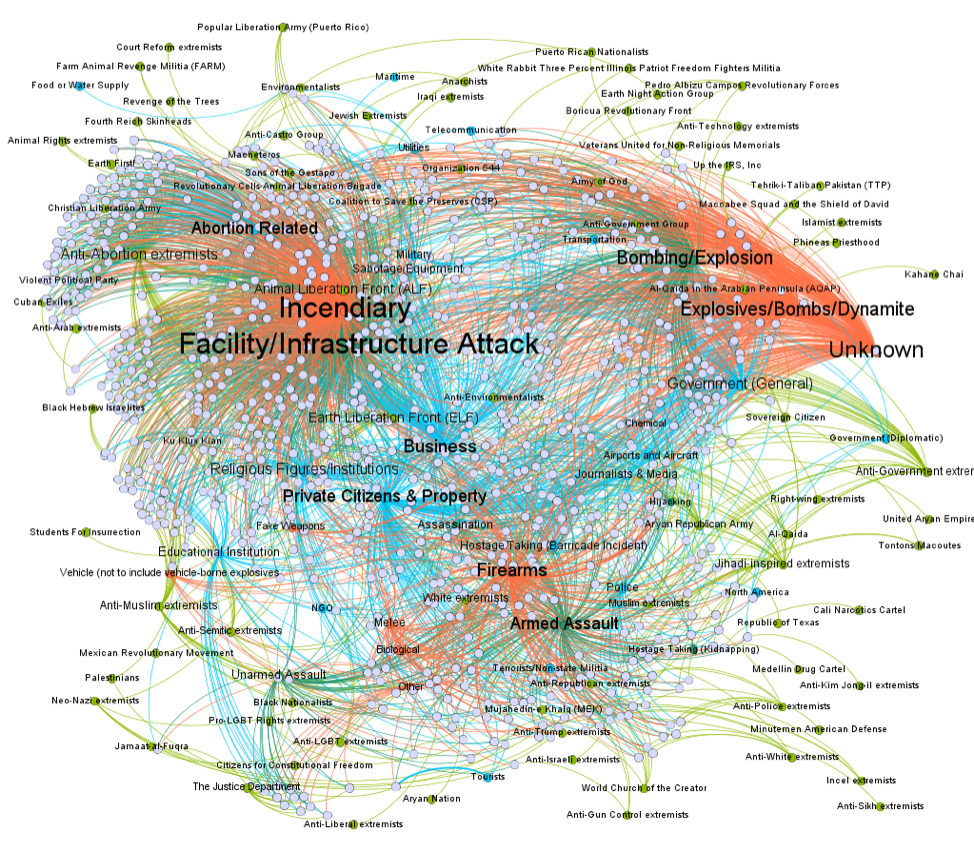}
	\caption{GTD Event Graph}
	\vspace{-2.5mm}
	\label{fig:basegtd}
\end{figure}

The tabular nature of these datasets makes it hard to bring out the relationships between the entities involved in the system. We propose to construct a different view of the tabular datasets in terms of a graph to analyze them. We focus on the affinity of a terrorist organization to other organizations, attack types, targets, and weapons. We start with creating some small networks from 1,000 incidences in GTD to demonstrate some of the possible node-type/edge-type combinations that may be useful in detecting potential Chemical, Biological, Radiological, Nuclear, and high yield Explosives (CBRNE) Weapon of Mass Destruction (WMD) activities. We filter events to include only US incidents past 1990. Figure \ref{fig:basegtd} shows a graph with nodes representing perpetrator groups, weapon types, attack types, targets, and events.

\begin{figure}
	\centering
	\includegraphics[width=280pt,height=300pt]{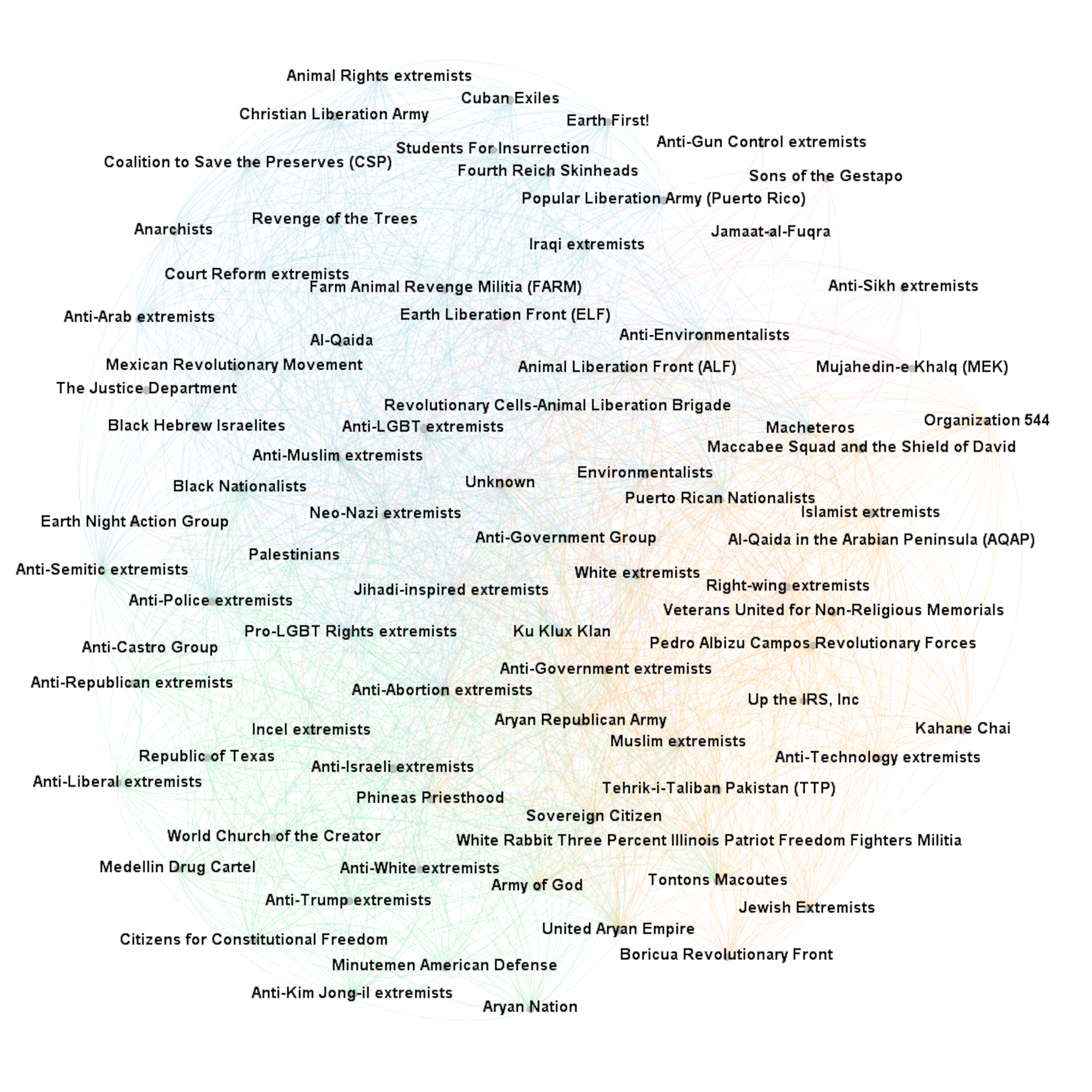}
	\caption{Weapon Profile of Terrorist organizations}
	\vspace{-2.5mm}
	\label{fig:orgweapon}
\end{figure}

\begin{figure}
	\centering
	\includegraphics[width=260pt,height=200pt]{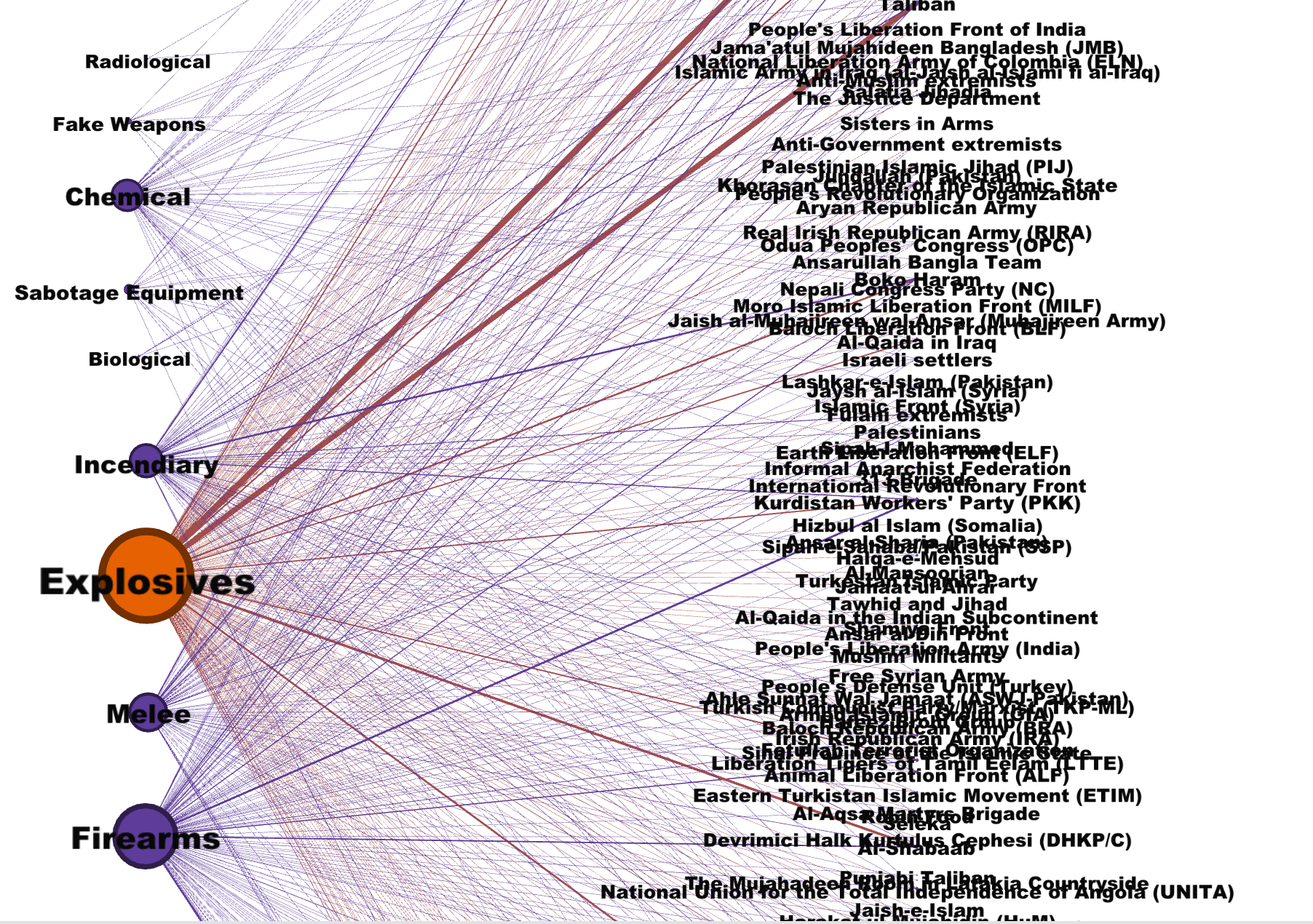}
	\caption{Bipartite View: Weapon Profile of Terrorist organizations}
	\vspace{-2.5mm}
	\label{fig:orgweaponbp}
\end{figure}

We are also interested in finding the weapon profile of a terrorist organization and finding groups of organizations using similar weapons. This is high value information which can be used to identify common sourcing of such weapons and to disrupt supply-chain of such organizations. Figure \ref{fig:orgweapon} shows a graph where nodes are the perpetrators, joined by the weapon types used in at least one incident. A bipartite graph view of the same network in figure \ref{fig:orgweaponbp} allows us to use graph connectivity of terror networks to compare their similarity in terms of modus operandi. 

\subsection{Temporal Graph Analysis} 
\begin{figure}
	\centering
	\includegraphics[width=230pt,height=120pt]{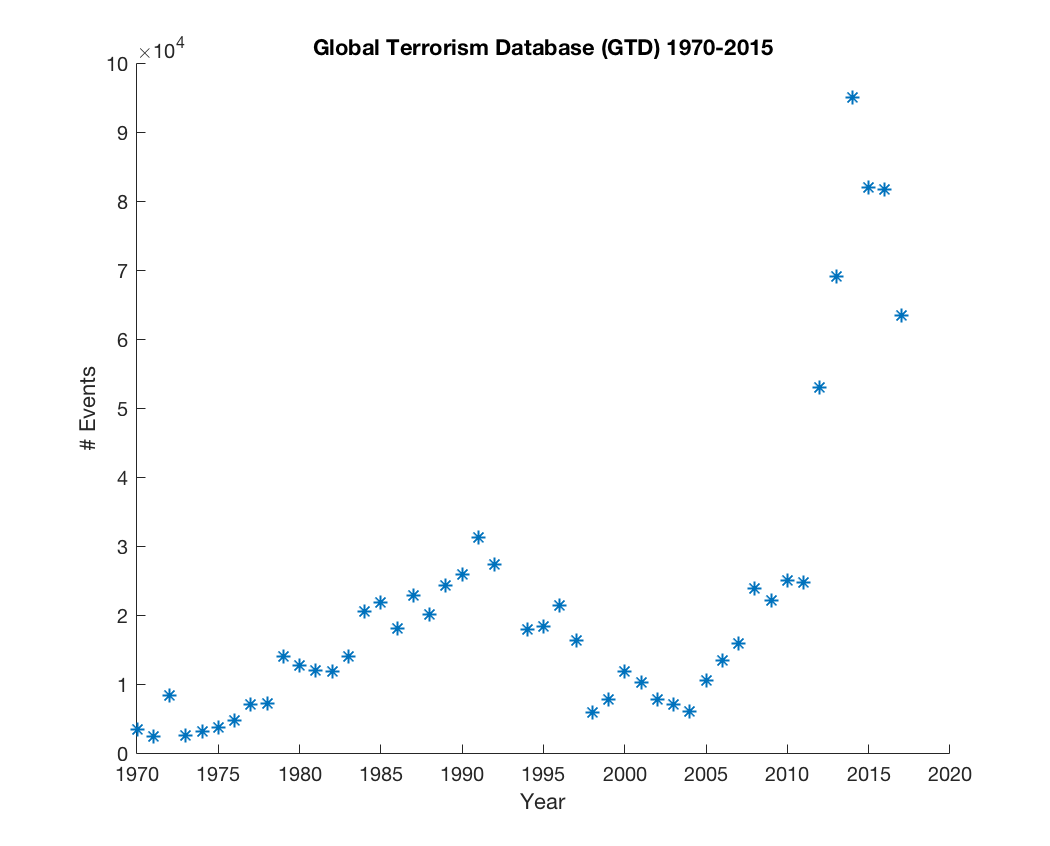}
	\caption{Yearly Frequency of GTD Events}
	\vspace{-2.5mm}
	\label{fig:yearlygtd}
\end{figure}

Temporal analysis of a complex system reveals many interesting and non-intuitive phenomena. Counting based measures such as Figure \ref{fig:yearlygtd} present limited information about the evolution of the network. Graphs are a powerful modeling tool to understand much more complex and latent properties of the system. 

\begin{figure}
	\centering
	\includegraphics[width=280pt,height=220pt]{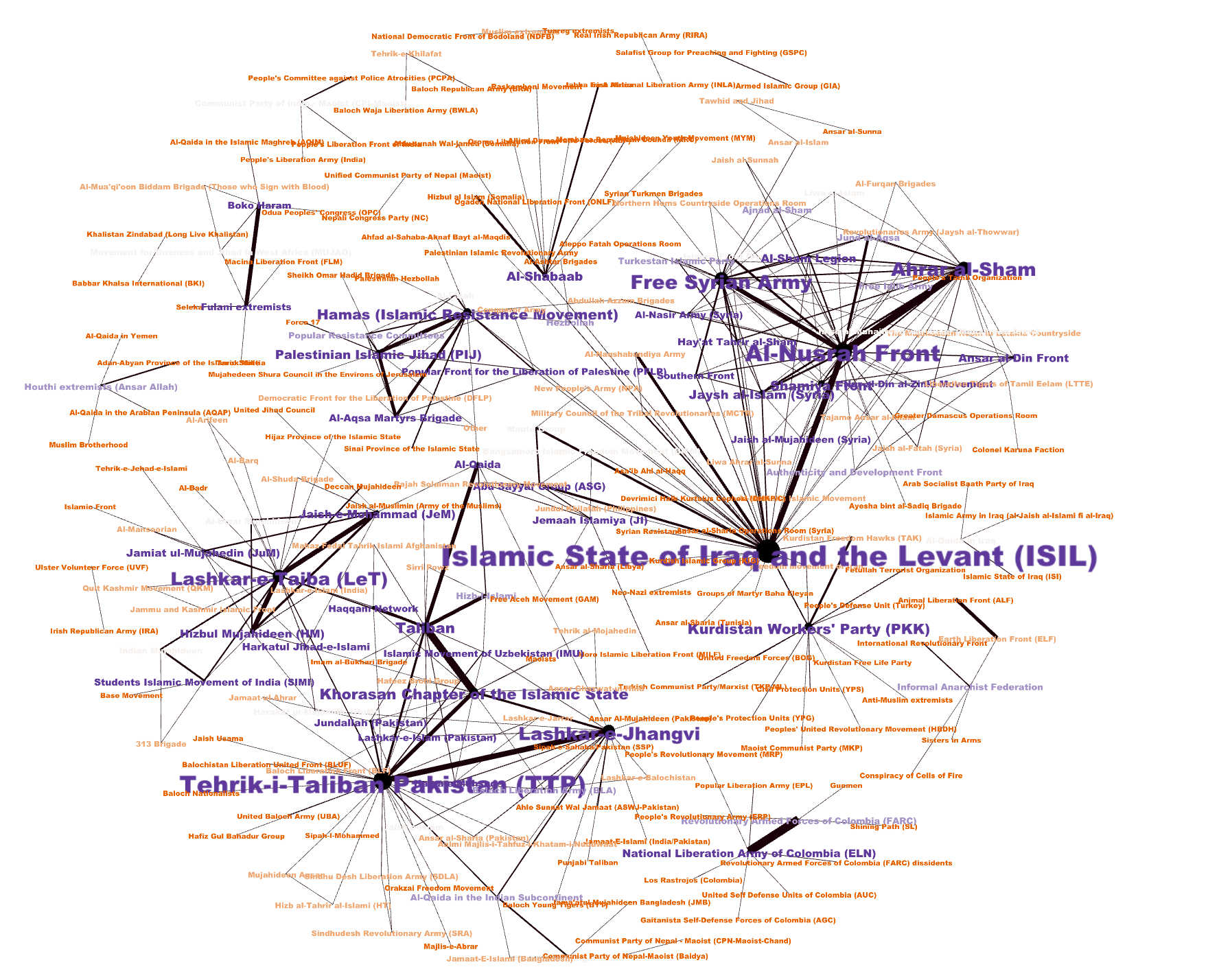}
	\caption{Communities of accomplice terrorist organization}
	\vspace{-2.5mm}
	\label{fig:assoOrg}
\end{figure}

\begin{figure}
	\centering
	\includegraphics[width=260pt,height=120pt]{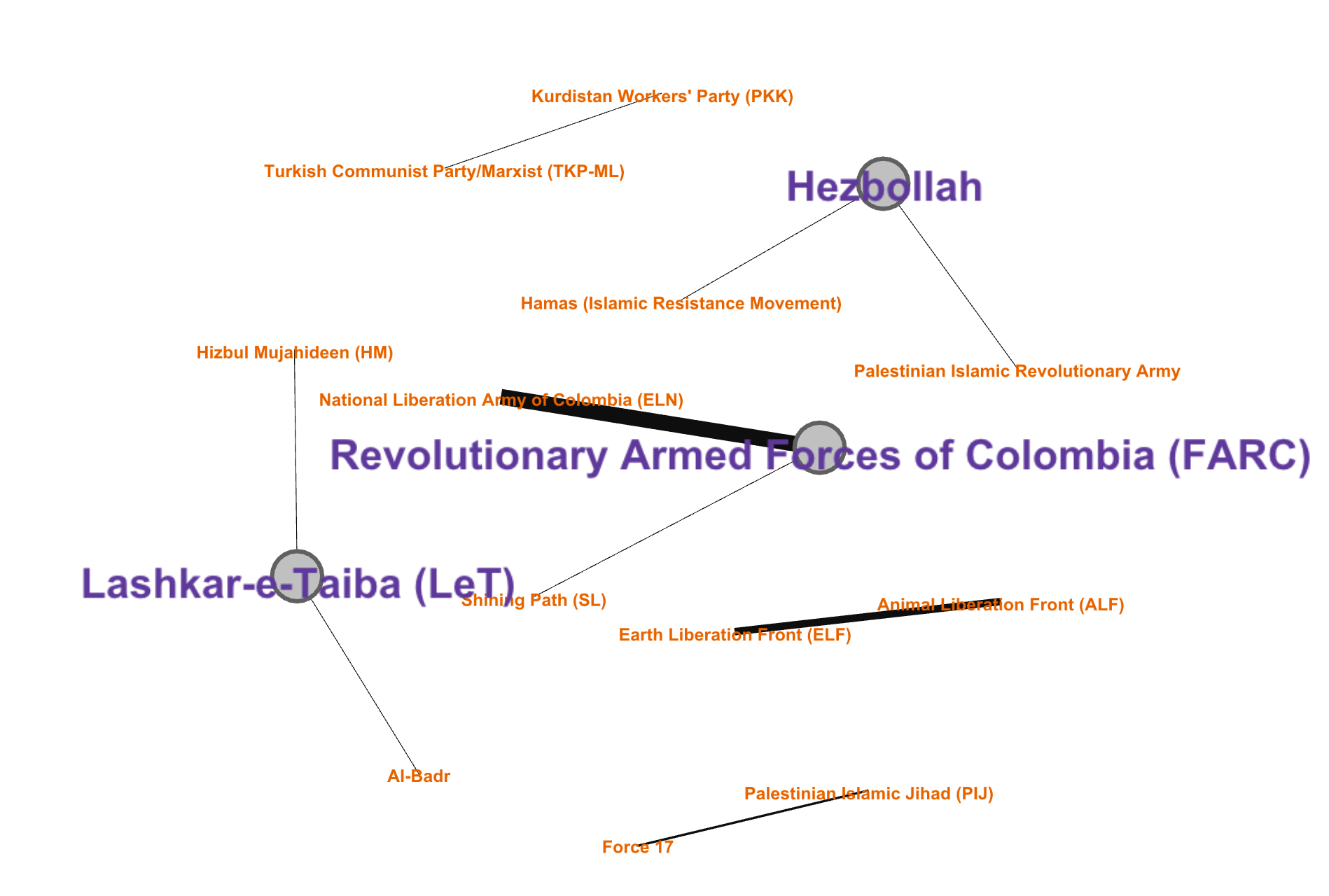}
	\caption{Communities of accomplice terrorist organization: 1990-2000}
	\vspace{-2.5mm}
	\label{fig:1990}
\end{figure}

\begin{figure}
	\centering
	\includegraphics[width=280pt,height=220pt]{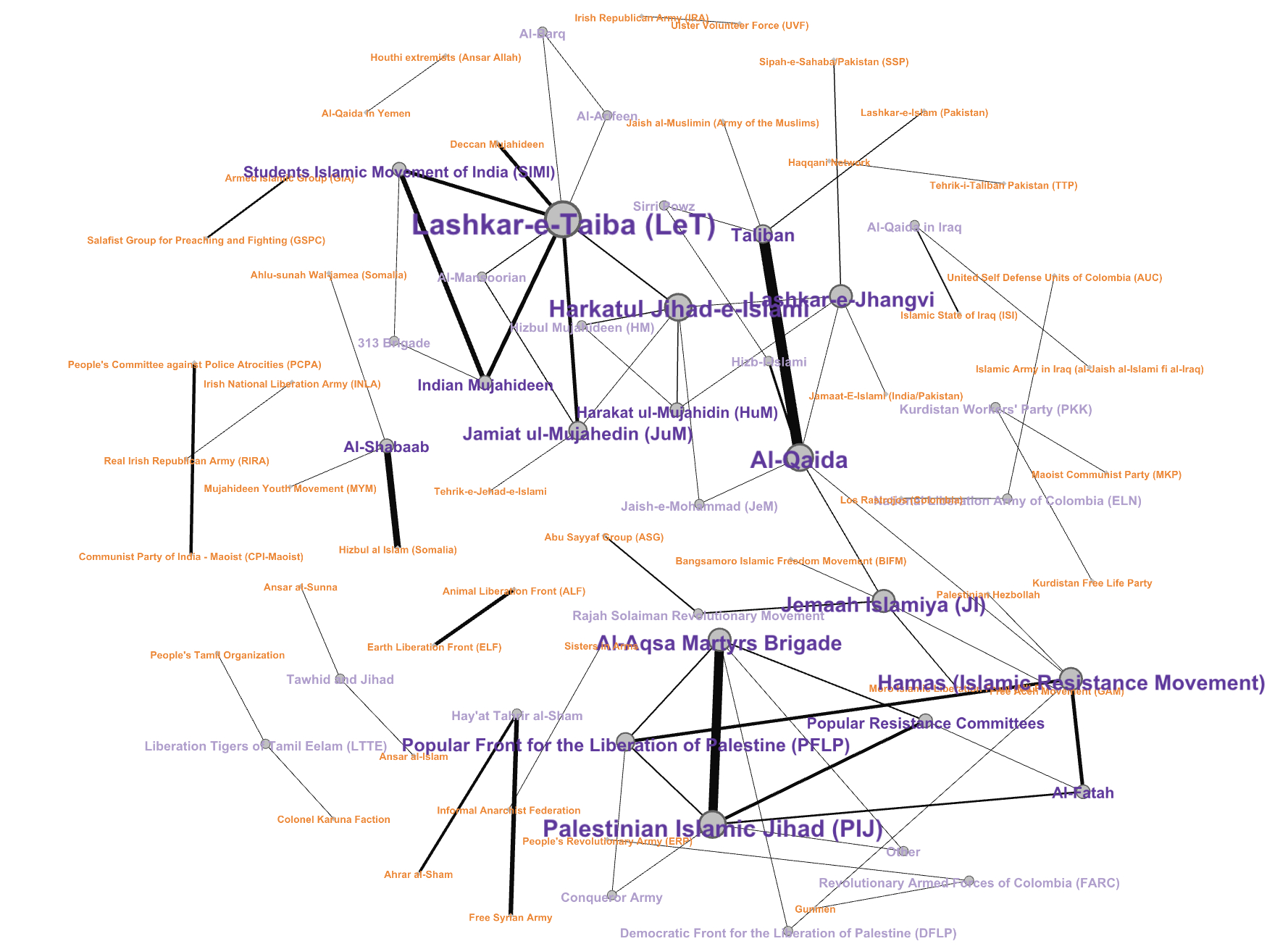}
	\caption{Communities of accomplice terrorist organization: 2001-2010}
	\vspace{-2.5mm}
	\label{fig:2000}
\end{figure}

\begin{figure}
	\centering
	\includegraphics[width=260pt,height=260pt]{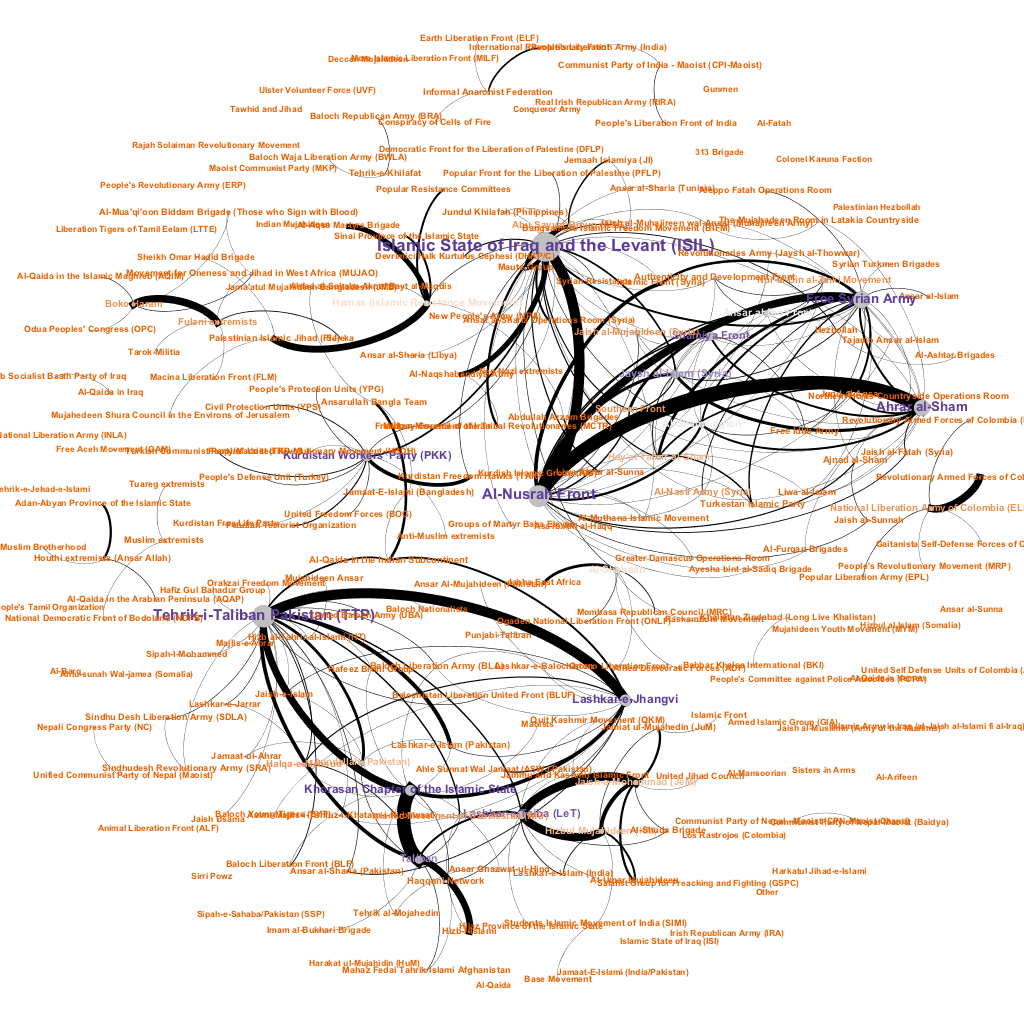}
	\caption{Communities of accomplice terrorist organization: 2011-2017}
	\vspace{-2.5mm}
	\label{fig:2011}
\end{figure}

We use small temporal dyads \cite{purohittemporal} to find terrorist organizations that are identified as multi-perpetrators of an event. In order to ensure consistency in the usage of group names for the database, the GTD database uses a standardized list of group names that has been established by project staff to serve as a reference for all subsequent entries \cite{lafree2007introducing}. Multiple perpetrator group attributions do not necessarily indicate that perpetrator groups collaborated to execute an attack. This could represent competing attributions, competing claims of responsibility, competing accusations, or a combination of these. We construct an association graph between main perpetrators of every event. For the sample GDT data beyond 1990, we get a clear indication of different communities of the terrorist organizations in figure \ref{fig:assoOrg}. The terrorist groups in a community may have co-ordinated some attacks together or claimed responsibilities for it. Figures \ref{fig:1990}, \ref{fig:2000}, and \ref{fig:2011} show a temporal shift in the active terrorist organization communities in terms of associated attack events. These types of higher order graph analyses of the terrorism network give much more insight into the operations of these groups, which is not available using lower order elements of the graph such as vertices and edges. We can clearly see which organization is central to facilitating such collaborations among smaller groups to conduct terrorist attacks. As shown in figure \ref{fig:2000}, Al-Qaida, Lashkar-e-Taiba (LeT), and PIJ were the hub of terrorist association between 1990-2000, but were overshadowed by ISIL and TTP after 2010. Similarly, edge thickness shows the strength of association between two groups. We also observe silos of operation for each terrorist group which gets bigger and denser over the course of time. LeT is such an example which can be seen growing its strength, association, and longevity from 1990 to 2017. The average degree of the association graph between 1990-2000 is 1.2, which increases to 1.97 during 2001-2010, and is maximum 2.5 during 2011-2017. Another interesting trend is observed before and after 2010, as modularity and average path length of the network decrease, indicating an increase in the number of isolated, scattered terror modules around the world.

\begin{figure}
	\centering
	\includegraphics[width=280pt,height=170pt]{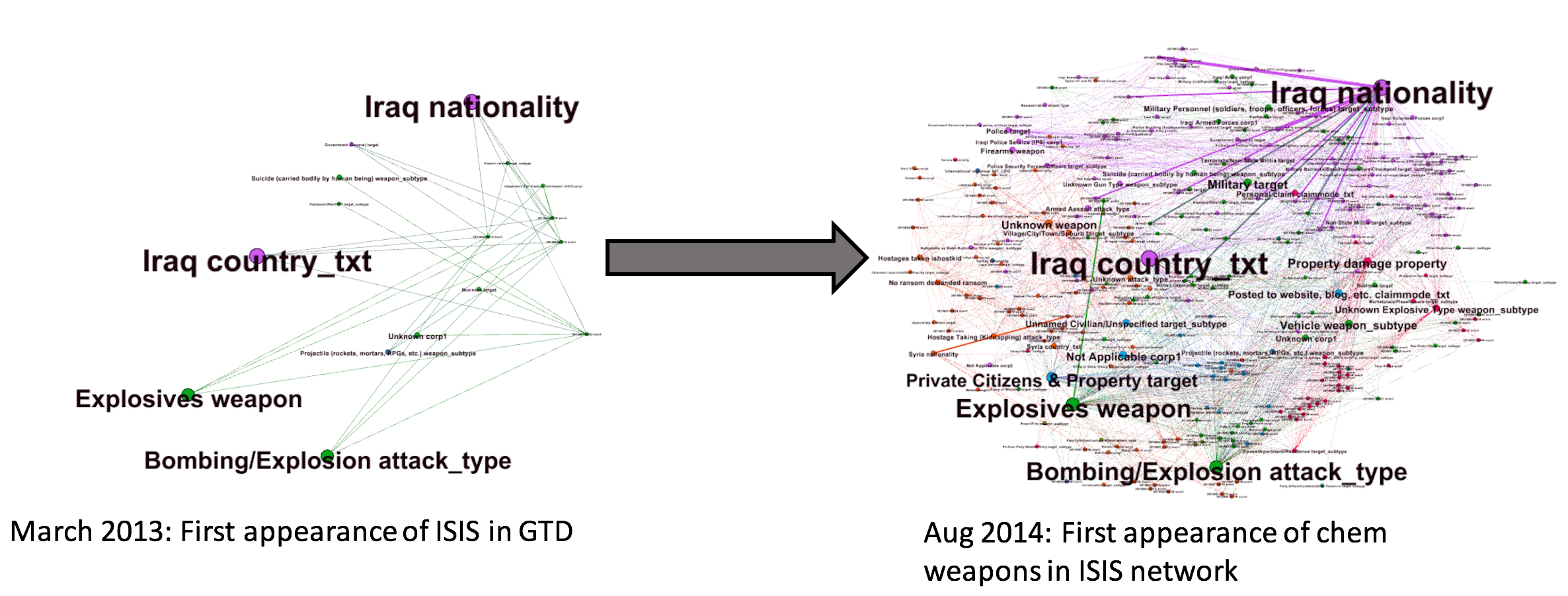}
	\caption{Temporal Network of ISIS}
	\vspace{-2.5mm}
	\label{fig:isisTempo}
\end{figure}

Similarly, a temporal network analysis allows us to examine the change in a group's behavior over time and events leading up to a significant event such as a chemical attack. Figure \ref{fig:isisTempo} shows a change in capabilities, target types, and activeness of ISIS over an 18 month period of time. 

\begin{figure}
	\centering
	\includegraphics[width=280pt,height=260pt]{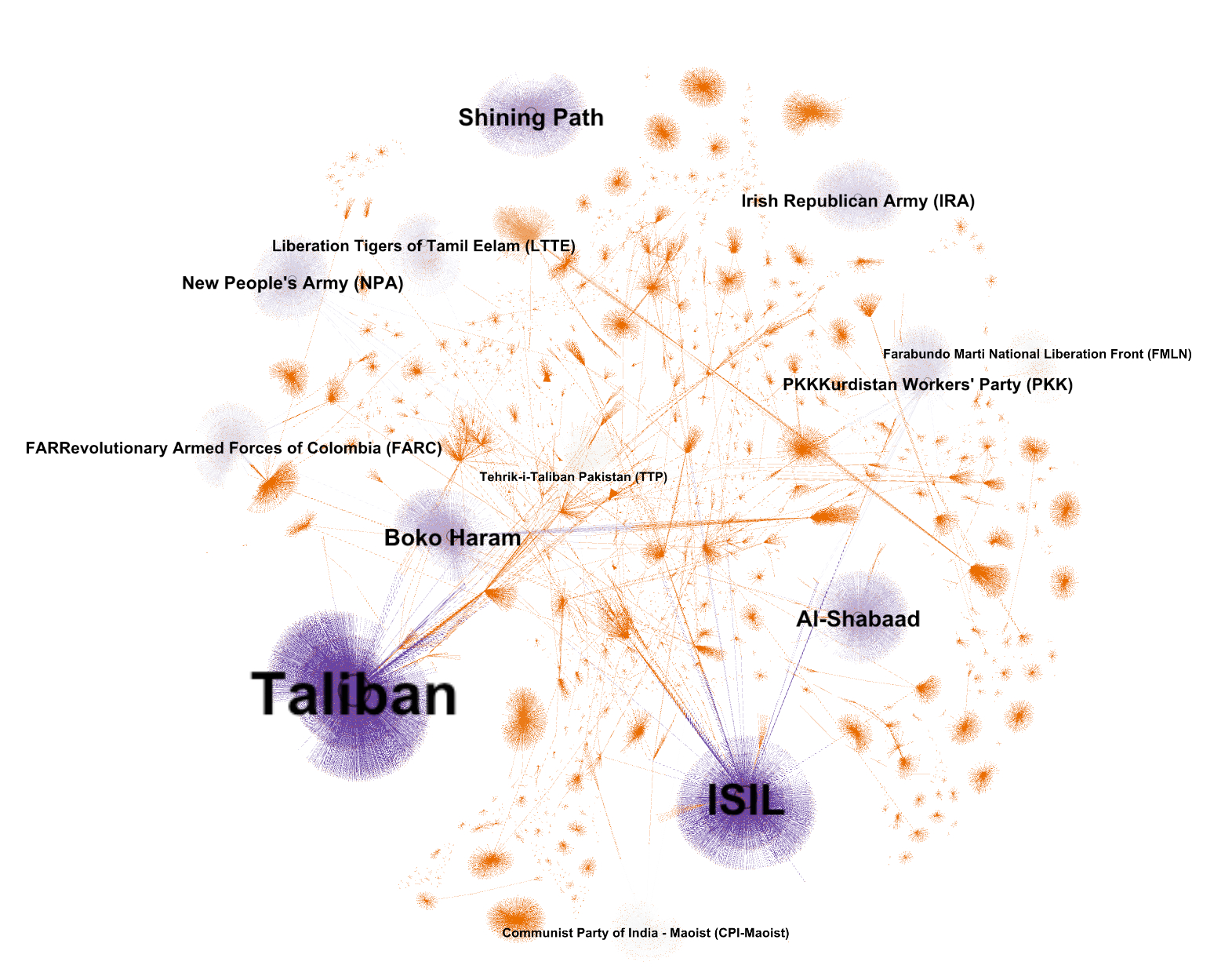}
	\caption{Top-k lethal terror organizations}
	\vspace{-2.5mm}
	\label{fig:lethality}
\end{figure}

Graphs also allow us to measure the impact of a terrorist organization in terms of number of casualties inflicted by an attack. We construct a higher order weighted graph between events and perpetrators with casualty count as an edge weight. We compute the lethality of a terror network from 1970 to 2017 using PageRank centrality measure. Figure \ref{fig:lethality} shows top-10 lethal terror organizations between 1970-2017.

\section*{Conclusion} 
International terrorism is a complex, ever-shifting threat and one of the biggest concerns of recent times. The global terrorism environment can be also described as a temporal network that evolves over time. We present a case study of analyzing the Global Terrorism Database (GTD) using graph based temporal analysis to reveal insights about different terror groups and their relationships with each other.

\bibliographystyle{IEEEtran}
\bibliography{gta3}

\end{document}